\documentstyle[preprint,eqsecnum,aps]{revtex}

\def\la{{\cal L}}
\def\na{{\cal N}}
\begin{document}
\draft
%\preprint{HEP/xxx-qed}
\title{Broken phase of the O$\bf (N)\,\phi^4$ model in light cone
 quantization and $\bf 1/N$ expansion\\}
\author{A. Borderies$^{\rm a}$, P. Grang\'e$^{\rm a,b}$ and
E. Werner$^{\rm c}$\\}
\address{$^{\rm a}$Centre de Recherches Nucl\'{e}aires,
IN2P3--CNRS/Universit\'{e} Louis Pasteur\\
F--67037 Strasbourg Cedex 2,  France\\}
\address{$^{\rm b}$Laboratoire de Physique Math\'{e}matique, \\
Universit\'{e} de Montpellier 2\\
F--34095 Montpellier Cedex 5,  France\\}
\address{$^{\rm c}$Instit\"ut f\"ur Theoretische Physik,\\
Universit\"{a}t Regensburg, Universit\"{a}tstra\ss e 31,\\
D--93053 Regensburg,  FRG\\}

\date{\today}
hep-th/9412104

\maketitle

\begin{abstract}
The solution of the O$(N)\,\phi^4$ scalar field theory in the broken
phase
is given in the framework of light cone quantization and a $1/N$
expansion.
It involves the successive building of operator solutions to the
equation
of motion and constraints including operator
zero modes of the fields which are the LC counterpart to the
equal time non trivial vacuum effects. The renormalization of the
procedure
is accomplished up to $2^{\rm nd}$ order in the $1/N$ expansion for
the equation
of motion and constraints. In addition the renormalization of the
divergent
contributions of the 2-point and 4-point functions is performed in a
covariant way.
The presence of a zero modes leads to genuine non perturbative
renormalization
features.
\end{abstract}

%\pacs{21.60.Cs, 21.60.Ev, 21.30.xy}

\section{Introduction}

In a recent paper [1] we have combined the Light-Front version of a
scalar
field theory possessing an O$(N)$ symmetry with an $1/N$ expansion.
For
the symmetric phase we have shown that this combination permits a
solution
of the field equations in terms of quadratures order by order in
$1/\sqrt N$.
This is in contrast to the equal-time quantized formulation where an
analogous solution is impossible. The physical origin of the
difference is the
triviality of the vacuum in the LC case as opposed to the increasing
complexity of the vacuum structure in the ET case, which forbids
solving the
equations of motion exactly beyond quadratic fluctuations.

In this paper we shall apply the techniques of ref. [1] to the
treatment of the
broken phase in which one of the components of the field operator
acquires
a non vanishing expectation value.

In section 2 we give the effective lagrangian as well as the
equations of motion and the
constraints. In section 3 we perform the expansion of the various
fields and
constraints in $1/\sqrt{N-1}$. In section 4 we describe the iterative
solution
order by order
in $\displaystyle{1\over\sqrt{N-1}}$ of the field equations including
the zero mode solution representative of the non perturbative
properties
of the system. In section~5 the renormalization of the procedure is
established showing again the importance of the determination of these
zero modes for the treatment of divergences and their appropriate
compensation. We conclude  in section~6 with an outlook to future
developments.

\section{Constrained algebra of the O$\bf (N)$ scalar $\bf\phi^4$
theory in the broken
phase}

The starting point is the effective lagrange density of the self
interacting scalar field with O$(N)$ internal symmetry. After
introduction of the auxiliary component
$\sigma$-field and integration over the first $N-1$ components, it is
the
remaining component $\phi^N(x)$ which will develop a non zero vacuum
expectation value. For LC dynamics $\la_{eff}(x)$ writes [2]:
\begin{eqnarray}
\la_{eff}(x)&=\displaystyle{1\over2}
\bigg\{&\partial^-\phi^N(x)\,\partial^+\phi^N(x)-
\left(\partial_\bot\phi^N(x)\right)^2-m^2\phi^N(x)\phi^N(x)\nonumber\\
&&-i\sigma(x)\phi^N(x)\phi^N(x)-{(N-1)\over g}\sigma^2(x)\nonumber\\
&&-(N-1)\,{\rm Tr}^y\,\ell n \left[(\partial^-\partial^+-
\partial_\bot^2
+m^2+i\sigma)\,\delta(x-y)\right]\bigg\}
\end{eqnarray}

We call $\phi(x)=\sqrt{N-1}\phi^N(x)$.
The conjugate momenta for the fields $\sigma(N)$ and $\phi(x)$
are obtained (in LC coordinates), as:
\begin{eqnarray}
\pi_\sigma(x)&\approx&0\\
\pi_\phi(x)&-&(N-1)\partial^+\phi(x)\approx0
\end{eqnarray}
Hence there exist two primary constraints:
\begin{eqnarray}\left\{\begin{array}{l}
\theta_\sigma(x)=\pi_\sigma(x)\approx0\\
\\
\theta_\phi(x)=\pi_\phi(x)-(N-1)\partial^+\phi(x)\approx0.
\end{array}\right.
\end{eqnarray}

Their consistency conditions yield two secondary constraints:
\begin{eqnarray}
\dot\theta_\sigma(x)&=&{1\over g}\sigma(x)-{i\over2}{\rm Tr}^y
\left[\left(\partial^+\partial^--
\partial_\bot^2+m^2+i\sigma(y)\right)^{-1}
\delta(x-y)\right]\nonumber\\
&&-{i\over2}\phi(x)\phi(x)\approx 0\\
\dot\theta_\phi^{P^-}(x^+,\roarrow{x_\bot})&=&P^-*\left[-(m^2+i\sigma-
\partial_\bot^2)\phi(x)\right]\approx 0.
\end{eqnarray}

$P^-$ is defined by
\begin{equation}
P^- *f(x)={1\over2L}\int_{-L}^L \,dx^-f(x),
\end{equation}
and projects onto the sector of operators independent of $x^-$, but
possibly dependent on $\roarrow{x_\bot}$ and $x^+$. The particle
sector is
obtained by projection with $Q^-=\openone-P^-$;
in order to obtain the vacuum sector mode of translational invariant
quantities one has to introduce another projector
$P^\bot P^-$ with $P^\bot=\displaystyle{1\over V_\bot}\int dx_\bot^{D-
2}$.
Obviously, the subspaces defined by $P^-$ and $P^\bot$ are not
orthogonals.

The dynamics of the fields $\sigma(x)$ and $\phi(x)$ is determined by
the
equation of motion, projected onto the $Q^-$-space,
\begin{equation}
Q^-*\left(\partial^-\partial^+-\partial_\bot^2
+m^2+i\sigma(x)\right)\,\phi^N(x)=0
\end{equation}
and the constraint (2.5), while the constraint (2.6) is just the
projection
of the equation of motion onto the $P^-$-space.

In the ET-formulation the equations of motion approach for the field
operators
is of no use if the
vacuum is non trivial. On the contrary in the LC-formulation  the
equations of
motions (E.M.) can in principle always be solved by taking appropriate
matrix elements between Fock space states, yielding an infinite set of
coupled equations.

A non perturbative solution of these equations being not available in
the
present case, one has to look for a small expansion parameter. As
shown
in ref.[3], such an expansion parameter is $1/N$, if the number of
field
components is sufficiently large.

\section{$\bf 1/N$ expansion}

Referring to the treatment of the symmetric phase in ref.[1] we
decompose
the $\sigma$-field in a constant background fiels $\sigma_0$ and its
complement $\tilde\sigma(x)$, which is developed in powers of
$1/\sqrt{N'}$
(where $N'=N-1$). This is easily understandable as the large number
$N$ making
a $1/N$ expansion possible is the large number of identical
particles, i.e.
$N-1$.
\begin{eqnarray}
\sigma(x)&=&\sigma_0+\tilde\sigma(x)\nonumber\\
&=&\sigma_0+{1\over\sqrt{N'}}\sigma_1(x)+{1\over N'}\sigma_2(x)+\cdots
\end{eqnarray}

In the same way we introduce a similar expansion for the $\phi$-field.
As can be seen from the form (2.1) for the effective lagrangian the
correct expansion for $\phi^N(x)$ is:
\begin{equation}
\phi^N(x)=\sqrt{N'}\omega_0+\phi_1(x)+{1\over\sqrt{N'}}\phi_2(x)+\cdot
\end{equation}

So, for $\phi(x)$ it is:
\begin{equation}
\phi(x)=\omega_0+{1\over\sqrt{N'}}\phi_1(x)+{1\over
N'}\phi_2(x)+\cdots
\end{equation}

Here, $\omega_0$ is a constant $C$-number. As we shall show below
$\phi_1(x)$ is entirely in the $Q^-$-space.

With the definitions (3.1) and (3.3) we arrive at expansions for the
constraints $\dot\theta_\sigma(x)$ and $\dot\theta_\phi^{P^-}(x^+,
\roarrow{x^\bot})$:
\begin{eqnarray}
\dot\theta_\sigma(x)=&-&\left[{\sigma_0\over g}+{i\over2}\Delta(x,x)+
{i\over2}\omega_0^2\right]\nonumber\\
&-&{1\over\sqrt{N'}}\left[{\sigma_1(x)\over
g}+{1\over2}\int\Delta(x,t)
\sigma_1(t)\Delta(t,x){d^Dt\over(2\pi)^D}+
i\omega_0\phi_1(x)\right]\nonumber\\
&-&{1\over N'}\left[{\sigma_2(x)\over g}+{1\over2}\int\Delta(x,t)
\sigma_2(t)\Delta(t,x){d^Dt\over(2\pi)^D}\right.\nonumber\\
&&\quad -{i\over2}\int\Delta(x,t)\sigma_1(t)\Delta(t,u)
\sigma_1(u)\Delta(u,x){d^Dt\over(2\pi)^D}{d^Du\over(2\pi)^D}\nonumber\
&&\quad +\left.{i\over2}\phi_1^2(x)+i
\omega_0\phi_2(x)\right]\nonumber\\
&-&{1\over N'\sqrt{N'}}\left[{\sigma_3(x)\over
g}+{1\over2}\int\Delta(x,t)
\sigma_3(t)\Delta(t,x){d^Dt\over(2\pi)^D}\right.\nonumber\\
&&\quad -{i\over2}\int\Delta(x,t)\sigma_1(t)\Delta(t,u)
\sigma_2(u)\Delta(u,x){d^Dt\over(2\pi)^D}{d^Du\over(2\pi)^D}
-{i\over2}(1\leftrightarrow2)\nonumber\\
&&\quad+{1\over2}\int\Delta(x,t)\sigma_1(t)\Delta(t,u)\sigma_1(u)\Delt
\sigma_1(v)\Delta(v,x){d^Dt\over(2\pi)^D}{d^Du\over(2\pi)^D}
{d^Dv\over(2\pi)^D}\nonumber\\
&&\quad+{i\over2}\left\{\phi_1(x)\phi_2(x)+(1\leftrightarrow2)\right\}
&&\quad+i\omega_0\phi_3(x)\Bigg]\nonumber\\
&+&O\left({1\over N^{'2}}\right)\approx 0,
\end{eqnarray}
which corresponds to $\dot\theta_\sigma(x)=-\dot\theta_\sigma^0(x)-
\displaystyle{1\over \sqrt{N'}}\dot\theta_\sigma^1(x)-
{1\over N'}\dot\theta_\sigma^2(x)+\cdots$
\begin{eqnarray}
\dot\theta_\phi^{P^-}(x^+,x^\bot)&=&(m^2+i\sigma_0)\omega_0\nonumber\\
&+&{1\over\sqrt{N'}}P^-*\left(-
\partial_\bot^2\phi_1(x)+i\omega_0\sigma_1(x)
\right)\nonumber\\
&+&{1\over N'}P^-*\left(i\sigma_1(x)\phi_1(x)
-\partial_\bot^2\phi_2(x)+i\omega_0\sigma_2(x)\right)\nonumber\\
&+&{1\over N'\sqrt{N'}}P^-
*\left(i\sigma_2(x)\phi_1(x)+i\sigma_1(x)\phi_2(x)
-\partial_\bot^2\phi_3(x)+i\omega_0\sigma_3(x)\right)\nonumber\\
&+&O\left({1\over N^{'2}}\right)\approx 0,
\end{eqnarray}
which corresponds to $\dot\theta_\phi^{P^-}(x^+,x^\bot)=
\dot\theta_\phi^0(x^+,x^\bot)+
\displaystyle{1\over \sqrt{N'}}\dot\theta_\phi^1(x^+,x^\bot)+
{1\over N'}\dot\theta_\phi^2(x^+,x^\bot)+\cdots$

These constraints have to be supplemented with the equation
of motion projected onto the $Q^-$-space (equation 2.8) which writes
 explicitely:
\begin{eqnarray}
&&Q^-*\left[(\partial^-\partial^+-
\partial^2_\bot)\,\phi_1(x)+i\omega_0
\sigma_1(x)\right]\nonumber\\
&+&{1\over \sqrt{N'}}
Q^-*\left[(\partial^-\partial^+-\partial^2_\bot)\,\phi_2(x)+
i\sigma_1(x)\phi_1(x)+i\omega_0\sigma_2(x)\right]\nonumber\\
&+&{1\over N'}
Q^-*\left[(\partial^-\partial^+-\partial^2_\bot)\,\phi_3(x)+
i\sigma_1(x)\phi_2(x)+i\sigma_2(x)\phi_1(x)+
i\omega_0\sigma_3(x)\right]\nonumber\\
&+&O\left({1\over N'\sqrt{N'}}\right)=0
\end{eqnarray}

Since the vacuum state $\vert0>$ is trivial, the operators in
eq.(3.4) to
eq.(3.6) can be determined by taking matrix elements between Fock
states and
the solutions can be written as sums of normal products of Fock
operators
(Haag series) with coefficients that have to be determined from the
pertinent  equation of motion or constraint.

\section{Iterative solution}

We shall now solve these constraints order by order in ${1\over
\sqrt{N'}}$.
In lowest order, the constraint $\dot\theta_\phi^0\approx 0$
sandwiched between the vacuum state yields the strong equality:
\hfill\break
$(m^2+i\sigma_0)\omega_0=0$, (since $\omega_0$ is constant) which
yields the
two possible phases of the theory:

\begin{itemize}
\item the symmetric phase treated in ref.[1], if we choose
$\omega_0=0$;
\item the broken phase, if we choose $\omega_0\not=0$ i.e.
\begin{equation}
m^2+i\sigma_0=0,
\end{equation}

which is an equation for $\sigma_0$ in terms of the
mass $(\sigma_0=im^2)$; the propagator $\Delta(x,x)$ becomes:
\begin{equation}
\Delta(x,x)={\rm Tr}^y [(\partial^+\partial^--\partial^2_\bot)]^{-1}
\delta(x-y)=\int^{\Lambda}{d^Dk_e\over(2\pi)^D}\,{1\over k_e^2},
\end{equation}
where $\Lambda$ symbolises the necessary regularisation.
\end{itemize}

If we put this result in the effective LC lagrangian density (2.1) we
get:
\begin{eqnarray}
\la_{eff}^{\rm LC}(x)={N-1\over2}&&\left\{\partial^-
\phi(x)\partial^+\phi(x)-
\left(\partial_\bot\phi(x)\right)^2-
i\tilde\sigma(x)\phi(x)\phi(x)\right.
\nonumber\\
&&\quad-\left.{\sigma^2(x)\over g}-{\rm Tr}^y
[(\partial^-\partial^+-\partial^2_\bot+i\tilde\sigma)
\delta(x-y)]\right\}
\end{eqnarray}

One sees that in the lowest order where $\sigma=\sigma_0$, there is
no mass
term for the field $\phi(x)$. In higher orders a mass term can appear
from
constant $C$-number parts of $\tilde\sigma$ which are at least of
order
${1\over\sqrt{N'}}$, which means that in the large $N$ limit the mass
term
would disappear. However, this reasoning is incomplete as we show
below.

In the broken phase one has an additional equation for the
determination of
$\omega_0$.

The constraint $\dot\theta_\sigma^0\approx0$ sandwiched between the
vacuum
state gives:
\begin{equation}
\omega_0^2=-{2\over g}\,m^2-\Delta(x,x).
\end{equation}

This equation has a solution with $\omega_0^2>0$ only for $D>2$.

\noindent
In next order one obtains
\begin{itemize}
\item  $\dot\theta_\phi^1(x^+,x^\bot)\approx0$ which results in
\begin{equation}
P^-*\sigma_1(x)=\sigma_1^{P^-}\approx0.
\end{equation}

\item  $\dot\theta_\sigma^1(x) \approx 0$ which yields:
\begin{equation}
\int\sigma_1(t)\,K(x,t)\,{d^Dt\over(2\pi)^D}+i\omega_0\,\phi_1(x)\appr
\end{equation}
with
$$K(x,t)={1\over g}\,\delta(x-t)+{1\over2}\,\Delta(x,t)\Delta(t,x).$$
\end{itemize}

{}From (4.5) one sees that $\sigma_1(x)$ belongs entirely to the
particle
sector.

To complete the solution one has to make use of the equation of motion
which to this order reads:
\begin{equation}
{1\over2}(\partial^-\partial^+-\partial^2_\bot)\phi_1(x)+i\sigma_1(x)
\omega_0=0.
\end{equation}

Using (4.6) to eliminate $\sigma_1(x)$ from (4.7) one gets:
\begin{eqnarray}
{1\over g}(\partial^-\partial^+-\partial^2_\bot)\phi_1(x)&+&
{1\over2}\left.\int\Delta(x,t)(\partial^-\partial^+-\partial^2_\bot)
\right\vert_t\phi_1(t)\Delta(t,x)\,{d^Dt\over(2\pi)^D}\nonumber\\
&+&\omega_0^2\phi_1(x)\approx0,
\end{eqnarray}
which shows that the $\phi_1$-field has now acquired a mass
$g\omega_0^2$.
In the LC formulation a Fock space expansion for the $\phi_1$ and
$\sigma_1$
fields is now useful. Therefore one can write:
\begin{eqnarray}
\phi_1(x)&=&\sum\limits_{n>0}\,C_n\,\left[a^+_ne^{-
ik_nx}+a_ne^{ik_nx}\right]\\
\sigma_1(x)&=&\sum\limits_{n>0}\,D_n\,\left[a^+_ne^{-ik_nx}+a_n
e^{ik_nx}\right].
\end{eqnarray}

Taking matrix elements of (4.6) and (4.7) between the vacuum and one
particle states yields the dispersion relation for the $k_n$'s:
\begin{equation}
\omega_0^2-k_n^2\left[{1\over g}+{1\over2}\int{d^Dq\over(2\pi)^D}\,
{1\over q^2(q+k_n)^2}\right]=0,
\end{equation}
and the relation between the $D_n$'s and the $C_n$'s:
$$D_n=i{C_n\,k_n^2\over\omega_0}.$$

%where the~~~~~field verifying eq.(3.14) is a free field and has
coefficients
%equal to

\vspace{1cm}
We consider now the order ${1\over N'}$ in the expansion. We need
$\dot\theta_\phi^2\approx0$, which is already projected in $P^-$-
space,
the $P^-$ and $Q^-$-spaces projection of $\dot\theta^2_\sigma\approx0$
and the $Q^-$-space projection of the equation of motion:
\begin{eqnarray}
\bullet\quad&& i\,P^-*\left(\sigma_1\phi_1(x)\right)+i\,\omega_0\,
\sigma_2^{P^-}(x^+,x^\bot)-\partial_\bot^2\phi_2^{P^-
}(x^+,x^\bot)\approx0\\
\bullet\quad&&{\sigma_2^{P^-}\over g}+{1\over2}\int
{dx^-dt^+d^{D-2}t^\bot\over(2\pi)^D}
\Delta(x,t)\sigma_2^{P^-}(t,x)\Delta(t,x)\nonumber\\
&&-{i\over2}P^-
*\int\Delta(x,t)\sigma_1(t)\Delta(t,u)\sigma_1(u)\Delta(u,x)
{d^Dt\over(2\pi)^D}\,{d^Du\over(2\pi)^D}\nonumber\\
&&+{i\over2}P^-*\left(\phi_1^2(x)\right)+i\omega_0\phi_2^{P^-}
(x^+,x^\bot)\approx0\\
\bullet\quad&&{1\over g}\sigma_2^{Q^-}+{1\over2}\int
{d^Dt\over(2\pi)^D}
\Delta(x,t)\sigma_2^{Q^-}(t)\Delta(t,x)\nonumber\\
&&-{i\over2}Q^-
*\int\Delta(x,t)\sigma_1(t)\Delta(t,u)\sigma_1(u)\Delta(u,x)
{d^Dt\over(2\pi)^D}\,{d^Du\over(2\pi)^D}\nonumber\\
&&+{i\over2}Q^-*\left(\phi_1^2(x)\right)+i\omega_0\phi_2^{Q^-}
(x)\approx0\\
\bullet\quad&&(\partial^-\partial^+-\partial_\bot^2)\phi_2^{Q^-}(x)+i
Q^-*(\sigma_1\cdot\phi_1)(x)+i\omega_0\sigma_2^{Q^-}(x)=0.
\end{eqnarray}
Here the notation is such that $Q^-*f(x)=f^{Q^-}(x)$.

By combining the two weak equations (4.12) and (4.13) one obtains a
weak
equation for $\phi_2^{P^-}(x^+,x^\bot)$:
\begin{eqnarray}
&-&{1\over g}\partial_\bot^2\phi_2^{P^-}(x^+,x^\bot)-{1\over2}\int
\Delta(x,t)\partial_\bot^2\phi_2^{P^-}(t^+,t^\bot)\Delta(t,x)
{dx^-d^{D-2}t^\bot dt^+\over(2\pi)^D}\nonumber\\
&&+\omega_0^2\phi_2^{P^-}(x^+,x^\bot) \nonumber\\
\approx&-&i\bigg[
{1\over g} \left(P^-*\sigma_1\phi_1\right)(x^+,x^\bot)+{1\over2}\int
\Delta(x,t)\left(P^-*\sigma_1\phi_1\right)(t^+,t^\bot)\Delta(t,x)
{dx^-d^{D-2}t^\bot dt^+\over(2\pi)^D}\bigg]\nonumber\\
&-&{\omega_0\over2}\left(P^-*(\phi_1^2\right)(x^+,x^\bot)\nonumber\\
&+&{\omega_0\over2}P^-
*\int\Delta(x,t)\sigma_1(t)\Delta(t,u)\sigma_1(u)
\Delta(u,x){d^Dt\over(2\pi)^D}{d^Du\over(2\pi)^D}.
\end{eqnarray}

In the same manner one obtains from (4.14) and (4.15)
\begin{eqnarray}
&&{1\over g}\left(\partial^-\partial^+-\partial_\bot^2\right)
\phi_2^{Q^-}(x)+\left.{1\over2}\int
\Delta(x,t)\left(\partial^-\partial^+-\partial_\bot^2\right)
\right\vert_{t}
\phi_2^{Q^-}(t)\Delta(t,x){d^Dt\over(2\pi)^D}
\nonumber\\
&&+\omega_0^2\phi_2^{Q^-}(x)\nonumber\\
\approx&-&i\bigg[
{1\over g} Q^-*\sigma_1\phi_1(x)+{1\over2}\int
\Delta(x,t)Q^-*(\sigma_1\phi_1(t))\Delta(t,x)
{dx^-d^{D-2}t^\bot dt^+\over(2\pi)^D}\bigg]\nonumber\\
&&-{\omega_0\over2}\left(Q^-*\phi_{(1)}^2\right)(x)\nonumber\\
&&+{\omega_0\over2}Q^-*
\int\Delta(x,t)\sigma_1(t)\Delta(t,u)\sigma_1(u)
\Delta(u,x){d^Dt\over(2\pi)^D}{d^Du\over(2\pi)^D}.
\end{eqnarray}

In the LC formulation these operator equations acquire significance
because
the action of the operators on Fock states is known.

Equation (4.16) is an inhomogenous linear integro-differential
equation for
$\phi_2^{P^-}(x^+,x^\bot)$. This equation
has no homogenous solution (i.e. solutions $\phi_2^{P^-}$ depending on
$x_\bot$ and $x^+$ but not on $x^-$). Therefore $\phi_2^{P^-}$ is
uniquely
determined by the inhomogenity which is made up of the already known
fields
$\sigma_1,\phi_1$ and $\omega_0$.

Its explicit form is:
\begin{eqnarray*}
\phi_2^{P^-}(x^+,x^\bot)&=&\sum\limits_{n^+,n^\bot,m^\bot}
\beta_{n^+,n^\bot,m^\bot}a_{n^+ n^\bot}^+a_{n^+m^\bot}\\
&&e^{-i{(k_n^{\bot^2}-k_m^{\bot^2})\over k_n^+}\,{x^+\over2}}\,
e^{-i(\vec k_n^\bot-\vec k_m^\bot)\,\vec x^\bot}.
\end{eqnarray*}

%The constant part $c$ leads to a shift of the vacuum expectation
value
%$\omega_0$ of the order ${1\over N'}$.
The sum on the right hand side can be
naturally decomposed in a global zero mode part (time and space
independent)
and a longitudinal zero mode (dependent on $\vec x_\bot$ and $x^+$).

The coefficient $\beta$ can be determined by taking matrix elements of
eq.(4.16) between the vacuum and
one particle Fock states.

The situation for the solution of eq.(4.17) is somewhat different
as there exist homogenous solutions of the operator on the l.h.s. of
the
equation which are identical to the solutions of equation (4.8) for
$\phi_1$. This means that just as in the case of the symmetric phase
$\phi_2^{Q^-}$ contains a one particle contribution which can be
lumped
together with $\phi_1$ and leads to an irrelevant change of the
normalisation.

Inspection of the r.h.s. shows that $\phi_2^{Q^-}$ contains a two-
particles
contribution which can be determined by taking matrix elements
between the
vacuum and the two-particles Fock sector. Also in this respect the
situation is analogous to the symmetric case.

This completes the solution $\phi_2^{P^-}$ and $\phi_2^{Q^-}$ and
hence
allows to find $\sigma_2^{P^-}$ and $\sigma_2^{Q^-}$ from equations
(4.12) and (4.15).

This procedure can in principle be continued by iteration. It such
allows
the solution of the equation of motion  order by order in
${1\over\sqrt{N'}}$.

Up to now the whole procedure was formal in the sense that it has
still to be
subject to the necessary renormalization. This can be done
order by order in ${1\over\sqrt{N'}}$ as will be shown in the next
section.

\section{Renormalization}
In the constraints (3.4) and (3.5) the expressions $\Delta(x,x)$ and
$\int
\Delta(x,t)\hat A(t)\Delta(t,x)d^Dt$, where the operator $\hat A$ can
be
either $\sigma_i(t)$ or ($\partial^-\partial^+-
\partial_\bot^2)\phi_i(t)$
exhibit divergences.

Their treatment can be done by a renormalization of the mass and the
coupling constant as will be shown below. Apart from these two
divergences
further divergent terms will appear in higher order, if fields
$\phi_2^{Q^-}$,
$\phi_3^{Q^-}\ldots $ obtained as solutions of the equation of motion
(3.6) in
higher orders, are used to calculate contributions to the two point
function
\break
$<0\vert T_+\phi^{Q^-}(x_1)\phi^{Q^-}(x_2)\vert 0>$. For instance
$\phi_2^{Q^-}$ will lead to one-loop-diagrams and $\phi_3^{Q^-}$ to
two-loop-diagrams; their divergent part have to be treated by field
renormalization. In addition, the necessity to perform a composite
field
renormalization will arise in order to make the vertex
$<0\vert \phi^{Q^-}(x_1)\phi(x_2)\sigma^{Q^-}(x_3)\vert 0>$ finite.
 There are no divergences coming from the
vacuum expectation values of products of fields operators in the
constraints
due to the normal product prescription (see the remark at the end
after
equation~(3.6)).

We now turn to the renormalization of the mass and coupling constant
mentioned
above:

$\quad$ We start with the lowest order term in eq.(3.4):
\begin{equation}
{1\over2}\omega_0^2+{m^2\over g}+{1\over2}\Delta(x,x)=0.
\end{equation}

The quantities $g_0$ and $\omega_0$ appear also in the equations of
motion
paralleling eq.(4.8) leading to the dispersion relation:
\begin{equation}
\omega_0^2-k_n^2\left[{1\over g}+{1\over2}k_n^{D-4}\int^{\Lambda/k_n}
{d^Dq\over(2\pi)^D}\,{1\over q^2(q+1)^2}\right]=0.
\end{equation}

Taking care of the divergent part of the integral
$I\left({\Lambda^2\over
k_n^2}\right)$ in equation (5.2) defines the renormalized coupling
$g_R$
at this order. If we renormalize by substraction at the energy scale
$\mu^2$ then, with $g=\lambda\mu^{4-D}$, the dimensionless coupling
$\lambda_R$
is obtained as:
\begin{equation}
{1\over\lambda_R(\mu^2)}={1\over\lambda}+{1\over2}I\left({\Lambda^2\ov
\mu^2}\right).
\end{equation}

In the broken symmetry phase one explores the domain of negative
squared
masses. Then setting $m^2=-m_R^2+\delta m^2$, using (5.3) in equation
(5.1),
one determines $\delta m^2$ as:
\begin{equation}
\delta m^2={1\over2}\lambda_R
\mu^2{\left[f\left({\Lambda^2\over\mu^2}\right)
-\left({m_R\over\mu}\right)^2I\left({\Lambda^2\over\mu^2}\right)\right
1-{\lambda_R\over2}I\left({\Lambda^2\over\mu^2}\right)},
\end{equation}
with
\begin{equation}
f\left({\Lambda^2\over\mu^2}\right)=\int^{\Lambda\over\mu}{d^Dx\over
x^2}.
\end{equation}

It is clearly seen from eqs(5.3) and (5.4) that for $D=4$ the mass
correction
$\delta m^2$ is governed essentially by the quadratically divergent
tadpole contribution $f\left({\Lambda^2\over\mu^2}\right)$, the
logarithmic
corrections coming from the replacement of $\lambda$ by $\lambda_R$
according to eq.(5.3).

Finally, eq.(5.1) reduces to:
\begin{equation}
{1\over2}g_R\omega_0^2=m_R^2
\end{equation}

Obviously this quantity is only half of the mass square which enters
in the
mass term of the equation of motion (4.8). This does not come as a
surprise
because eq.(4.8) being already of higher order in $1/\sqrt{N'}$
contains
the effect of $\sigma_1$ (in terms of $\phi_1$). The mass term in this
equation results from the local part of the interaction between the
fields
$\phi_1$ and $\sigma_1$. Including the non local part of this
interaction
leads to the dispersion relation (5.2).
Its solution yields the physical mass $M$ i.e. the mass corresponding
to the
pole of the single particle propagator.

The renormalization which we have performed so far exhausts
the divergence
appearing in the equations of motion and constraints.

The renormalization program is not yet completed as one has also to
take
care of divergences in correlation functions. A priori it is not
obvious
that the procedure for performing this task is the standard one. The
main
reason lies in the subtelties related to the infrared properties of
the
LC version. It implies that in all integrations over internal momenta
one
has to cut out the contributions coming from the neighbourhood of
vanishing
longitudinal momentum (which can be done in a well-defined manner in
the
discretized version).Diverging contributions come from one- and two-
loop
diagrams with local vertices,i.e. from the pieces in the expansions
of the
fields which are proportional to~$\phi_1^2$~and~$\phi_1^3$ [4]. In
the two-point
functions these parts give rise to the diagrams:

\begin{center}
\unitlength=1.00mm
\linethickness{0.4pt}
\begin{picture}(120.00,42.00)
\put(60.00,20.00){\circle{14.00}}
\put(67.00,20.00){\circle*{2.00}}
\put(53.00,20.00){\circle*{2.00}}
\put(9.00,21.00){\makebox(0,0)[cc]{{\Large $\Sigma :$}}}
\put(53.00,20.00){\dashbox{2.00}(0.00,22.00)[cc]{}}
\put(67.00,20.00){\dashbox{2.00}(0.00,22.00)[cc]{}}
\put(49.00,41.00){\makebox(0,0)[cc]{{\Large $\omega$}}}
\put(71.00,41.00){\makebox(0,0)[cc]{{\Large $\omega$}}}
\put(67.00,20.00){\line(1,0){33.00}}
\put(53.00,20.00){\line(-1,0){33.00}}
\end{picture}

\vspace{1cm}

\unitlength=1.00mm
\linethickness{0.4pt}
\begin{picture}(120.00,24.00)
\put(60.00,17.00){\circle{14.00}}
\put(67.00,17.00){\circle*{2.00}}
\put(53.00,17.00){\circle*{2.00}}
\put(9.00,18.00){\makebox(0,0)[cc]{{\Large $\Sigma^{\prime} :$}}}
\put(20.00,17.00){\line(1,0){80.00}}
\end{picture}
\end{center}
%\null\vspace{3.0cm}
which we call~$\Sigma$~and~$\Sigma'$~respectively.

{}From the fully discretized version we go over to a mixed one where
only the
longitudinal momenta are discretized whereas~$\vec p_\bot$ is a
continuous two-dimensional vector.
This is possible since only for the longitudinal
component~$k^+=0$~plays a special role.

The self energie~$\Sigma$~is then:
\begin{eqnarray*}
\Sigma&\sim&{1\over \na}g_R^2{\omega_0^2\over2}\sum\limits_{1}^{\na-1}
{1\over x_n(1-x_n)}\\
&\cdot&\int d^{D-2} p_\bot{1\over K^+K^--{M^2+\vec p_\bot^2\over x_n}-
{M^2+(K_\bot-\vec p_\bot)^2\over1-x_n}+i\varepsilon}.
\end{eqnarray*}
Here we use the notations:
$$K^+=\pi{\na\over L}\,;\qquad k_n^+=\pi{n\over L}\,;\qquad
x_n={n\over \na}$$
$K^-$ is written in units of $L/\pi$.
Clearly $\Sigma$ does not depend explicitely on $L$. The sum over $n$
 excludes the
alleged problematic IR divergence and goes over a finite number of
terms.
Hence the divergence of the diagram comes only from the integration
over
transverse momenta. If there is no limitation on the integration over
$\vec p_\bot$
the integral can be rewritten as (at one place one has to complete a
square
and shift $\vec p_\bot$, which is possible only if there is no
cutoff, since
otherwise the shifted cutoff would depend on $x_n$~and $\vec K_\bot$):
\begin{equation}
\Sigma \sim {1\over\na} g_R^2{\omega_0^2\over2}\sum\limits_{1}^{\na-1}
\int d^{D-2} p_\bot{1\over K^2x_n(1-x_n)-
(p_\bot^2+M^2)+i\varepsilon}.
\end{equation}
As explained above the integral over $\vec p_\bot$ can not be
regularized by a cutoff
on $\vec p_\bot^{\,2}$. Therefore we perform in the following a
dimensional regularization
with a minimal subtraction. Technically we proceed by the
Schwinger-representation for the integrand in eq.(5.7)[5]:
$${1\over K^2x_n(1-x_n)-(p_\bot^2+M^2)+i\varepsilon}={1\over
i}\int_0^\infty
e^{i\alpha[K^2x_n(1-x_n)-p_\bot^2-M^2+i\varepsilon]}d\alpha.$$

The~$\vec p_\bot$-integration in dimension $D-2$ introduces a factor
$\displaystyle{1\over \alpha{D\over 2}-1}$ into the
integrand which leads to the necessity of a lower cutoff
$\displaystyle{1\over
\Lambda^2}$
for the
$\alpha$-integration. $D$ is chosen as $D=4-\varepsilon$.
The $\alpha$-integration leads then to an
incomplete gamma function $\Gamma\left(\displaystyle{\varepsilon\over
2},\;
i{M^2-x_n(1-x_n)K^2\over\Lambda^2}\right)$
which can be evaluated employing
$$\Gamma\left({\varepsilon\over 2},x\right)=
\Gamma\left({\varepsilon\over 2}\right)-\sum\limits_{n=0}^\infty
{(-1)^nx^{{\varepsilon\over 2}+n}\over n!\,
\left({\varepsilon\over 2}+n\right)}.$$
Using
$$x^{{\varepsilon\over2}}=\exp ({\varepsilon\over2}\,\ell n\,x)\quad
\to\quad 1+
{\varepsilon\over2}\,\ell n\,x,$$
one sees that the~divergence in $\varepsilon/2$ of the gamma function
is
compensated by the $n=0$
contribution of the sum over $n$. One is left with
$$\Sigma \sim- {1\over 2} \omega_0^2 g_R^2\,\sum\limits_{n=1}^{\na-1}
\ell n\left[
{M^2-x_n(1-x_n)K^2\over\Lambda^2}\right].$$

Performing a subtraction at the renormalization point $K^2=\mu^2$ one
 decomposes $\sigma$ into
\begin{eqnarray*}
\Sigma&\sim&{\omega_0^2\over2}g_R^2\,\sum\limits_{n=1}^{\na-1}\ell n\,
{M^2-x_n(1-x_n)\mu^2\over M^2-x_n(1-x_n)K^2}\\
&-&{\omega_0^2\over2}g_R^2\,\sum\limits_{n=1}^{\na-1}\ell n\,
{M^2-x_n(1-x_n)\mu^2\over \Lambda^2},
\end{eqnarray*}
i.e. into a part remaining finite upon $\Lambda^2\to\infty$
and a logarithmicaly divergent one.

The sum over $n$
remains to be done. Analytically this is most easily done
in the limit $L\to\infty$. In order to keep $K^+$~fixed one has to
send $L$
 and~$\na$~to infinity keeping~$\displaystyle{\na\over L}$~fixed.
The inclusion of $x_n=0$ and $x_n=1$ doesn't change the value of the
sum
in the limit of $L\to\infty$. Therefore, replacing the sum by
integral, the
limits of integration can be taken 0 and 1.
One obtains in this way:
$$\lim\limits_{\na\to\infty} {1\over \na} \sum\limits_{n=1}^{\na-1}
F(x_n)=\int_0^1 dx\,F(x).$$

The integral over $x$ can be done by changing $x$ to
$x=\displaystyle{1\over2}+z$. In this way one
finally arrives at

\samepage
\begin{eqnarray}
\Sigma&\sim&{\omega_0^2\over2}g_R^2\,\ell n\,
{M^2\over M^2+{\mu^2-K^2\over4}}\nonumber\\
&+&2\omega_0^2\,g_R^2\sqrt{{M^2\over \mu^2}-{1\over4}}\;{\rm arctg}\;
 {1\over2\sqrt{{M^2\over \mu^2}-{1\over4}}}\nonumber\\
&-&2\omega_0^2\,g_R^2\sqrt{{M^2\over K^2}-{1\over4}}\;{\rm arctg}\;
{1\over2\sqrt{{M^2\over K^2}-{1\over4}}}\nonumber\\
&+&{\omega_0^2\over2}\,g_R^2\,\ell n
{M^2+{\mu^2-K^2\over4}\over \Lambda^2}.
\end{eqnarray}

The last divergent term can be compensated by a covariant counterterm
which
has to be added to the Lagrangian while the first three finite terms
lead to a
finite shift of the pole mass. The self energy $\Sigma$ is a typical
nonperturbative effect signaled by the factor $\omega_0^2$.
We therefore see here a nonperturbative
renormalization at work.

At the same order in $1/N$~a coupling constant renormalization has to
be
performed,since the above one loop intermediate state also
contributes to the
four point function according to the diagram

\begin{center}
\unitlength=1.00mm
\linethickness{0.4pt}
\begin{picture}(120.00,30.33)
\put(60.00,17.00){\circle{14.00}}
\put(67.00,17.00){\circle*{2.00}}
\put(53.00,17.00){\circle*{2.00}}
\put(67.00,17.00){\line(5,2){33.00}}
\put(67.00,17.00){\line(5,-2){33.00}}
\put(53.00,17.00){\line(-5,2){33.00}}
\put(53.00,17.00){\line(-5,-2){33.00}}
\end{picture}
\end{center}
%\vspace{2.5cm}

In constrast to the case of~ $\Sigma$~this coupling constant
renormalization is
perturbative in the sense that it has to be done also if $\omega_0=0$.
The diagram leads (apart from pre-factors) to the same analytic
expression
as~$\Sigma$~and
renormalizes the coupling~$g_R$ which was introduced upon the
renormalization
of constraints and equations of motion.

The two-loop diagram
$\Sigma'$
is also perturbative in the sense just mentioned. It can be evaluated
with
techniques quite analogeous to the case of~$\Sigma$, the essential
difference
being an additional two dimensional transverse momentum integration.
This
changes the dimensional argument~$\varepsilon/2$
which appeared in the incomplete gamma
function in~$\Sigma$~into~$\varepsilon-1$.
The result of the transverse integration is then
\begin{eqnarray}
\Sigma'&\sim&g_R^2 {1\over \na^2}\sum\limits_{n=1}^{\na-1}
\sum\limits_{m=1}^{\na-n-1}{1\over x_ny_m(1-x_n-y_m)}\nonumber\\
&&\left\{-\Lambda^2+A_{mn}\,\ell n\,{A_{mn}\over\Lambda^2}+A_{mn}(1-
\gamma)+
\cdots\right\}\end{eqnarray}
$\gamma$~is Euler's constant, and
$$A_{mn}=M^2\left(x_n(1-x_n)+y_m(1-x_n-y_m)\right)-K^2x_ny_m(1-x_n-
y_m).$$

A substraction at the renormalization point $K^2=\mu^2$ eliminates
the quadratically divergent term $\sim \Lambda^2$ but not the
logarithmic
divergence (as in the ET case). The latter one is taken care of by a
field renormalization via [4] $$\Sigma'_R=Z\Sigma'$$
Contrary to the previous one-loop case the sums over $n$ and $m$ do
not
lead to integrals that can be performed analytically. Moreover, due
to the
appearence of $x_n y_m (1 - x_n - y_m)$ in the denominator the sums
over $n$ and $m$ can not be extended to $n$ or/and $m=0$. Therefore
the sums
have to be carried out numerically. It can be seen that in the limit
$L \rightarrow \infty$ with ${\cal N}/L$ fixed the double sum is
proportional to ${\cal N}^2$; therefore in the large $L$ limit the
result
for $\Sigma'$ is independent of the external momentum
$K^+$. The finite
parts depend only on $m^2$ and $K^2$ while the required counterterms
depend in addition on the cutoff $\Lambda^2$ which has a covariant
meaning.

\vspace{0.4cm}
It is thus seen that the decomposition of the fields $\Phi^N$ and
$\sigma$ into a zero-mode part and a particle field with nonvanishing
longitudinal momentum allows a perfectly covariant renormalization. In
addition to the perturbative pieces which have their counterparts in
the
ET treatment there is also a nonperturbative piece yielding a nice
example for a nonperturbative renormalization. A corresponding
treatment
in the ET formulation would be quite impossible due to the unknown
structure of the groundstate at the considered order in $1/N$.

\newpage
\section{Conclusion}

The alleged simplicity of the ground state vacuum in light cone
quantization has long been the source of much suspiscion in its
capacity to
cope with essential features of non perturbative physics, for they
are known
to manifest themselves, in the conventional equal time quantization
scheme,
through non trivial vacuum effects. Taking seriously the singular
nature
of LC Lagrangians in the LC metric led to the understanding of the
specific
role of the zero modes of the field operators in describing non
perturbative
effects in this scheme. The operator valued zero modes have to be
determined
from non linear operator equations, which, in the absence of a small
expansion parameters, can only be solved approximatively. However for
theories with internal $N$ fold symmetry the situation is different.
Considering the O$(N)$ $\phi_{3+1}^4$ scalar field theory in the
symmetric
phase we have shown previously that its LC-version combined with an
appropriate $1/\sqrt N$ expansion is solvable in terms of quadrature
order
by order in $1/\sqrt N$. Here we focussed on the solution of the
equation of
motion and constraints in the broken phase. Although less
straigthforward
than in the symmetric phase, the solution follows the same
progression,
the successive orders being now in $1/\sqrt{N-1}$.

In both phases the essential features of the solutions can be
summarized as
follows: at every step in the expansion the solution for the composite
field $\sigma(x)$ (and $\phi^N(x)$ in the broken phase) has
contributions
in the sector of translational invariant operators (vacuum sector $P^-
$) and
its complement (particle sector $Q^-$). At the stage where in the ET
formulation
a non trivial vacuum appears an operator valued zero mode in the
vacuum
sectors shows up in the LC version.

The renormalization procedure is finally carried through. Whereas the
ultraviolet problem turns out to be standard, a genuine non
perturbative
aspect of renormalization shows up: it is only after the
determination of
the zero modes of the field operators that certain divergence can be
isolated
and compensated by appropriate counter terms. This decomposition is
also
essential in order to arrive at a covariant renormalization.
It is argued that the procedure can be continued to
higher orders but a general proof is still needed.

Another important issue which has to be understood
concerns the handling of non trivial topologies. In the framework of
LCQFT
this calls for the examination of models
where instantons are known at the classical level ($CP^{N-1}\ldots$)
[3]
and of their fates in the
whole Dirac-Bergmann algorithm. Clearly this is an essential step
before attacking the strong interaction domain of QCD.

\acknowledgments
We thank  S. Pinsky, H.C. Pauli and T. Heinzl for discussions. This
work
has been completed under NATO grant n$^\circ$ CRG 92.0472.

\end{document}